\documentclass[prl,showpacs,amsmath,amssymb,twocolumn, 9pt]{revtex4}

\usepackage{amsthm}
\usepackage{dcolumn}
\usepackage{amsmath}
\usepackage{bm}
\usepackage{graphicx}
\usepackage{xcolor}
\usepackage{ulem}

\newtheorem{prop}{Proposition}

\begin{document}

\newtheorem{corollary}{Corollary}
\newtheorem{definition}{Definition}
\newtheorem{example}{Example}
\newtheorem{lemma}{Lemma}
\newtheorem{proposition}{Proposition}
\newtheorem{theorem}{Theorem}
\newtheorem{fact}{Fact}
\newtheorem{property}{Property}
\newcommand{\bra}[1]{\langle #1|}
\newcommand{\ket}[1]{|#1\rangle}
\newcommand{\braket}[3]{\langle #1|#2|#3\rangle}
\newcommand{\ip}[2]{\langle #1|#2\rangle}
\newcommand{\op}[2]{|#1\rangle \langle #2|}

\newcommand{\tr}{{\rm tr}}
\newcommand {\E } {{\mathcal{E}}}
\newcommand {\F } {{\mathcal{F}}}
\newcommand {\diag } {{\rm diag}}

\title{Four Locally Indistinguishable Ququad-Ququad Orthogonal Maximally Entangled States}
\author{Nengkun Yu}
\email{nengkunyu@gmail.com}
\author{Runyao Duan}
\email{runyao.duan@uts.edu.au}
\author{Mingsheng Ying}
\email{mying@it.uts.edu.au}

\affiliation{State Key Laboratory of Intelligent Technology and Systems, Tsinghua National Laboratory\protect\\
for Information Science and Technology, Department of Computer Science and Technology,\protect\\
Tsinghua University, Beijing 100084, China}
\affiliation{Center for Quantum Computation and Intelligent Systems (QCIS), Faculty of
Engineering and Information Technology, University of Technology,
Sydney, NSW 2007, Australia}
\date{\today}
\begin{abstract}
We explicitly exhibit a set of four ququad-ququad orthogonal maximally entangled states that cannot be perfectly distinguished by means of local operations and classical communication. Before our work, it was unknown whether there is a set of $d$ locally indistinguishable $d\otimes d$ orthogonal maximally entangled states for some positive integer $d$. We further show that a $2\otimes 2$ maximally entangled state can be used to locally distinguish this set of states without being consumed, thus demonstrate a novel phenomenon of ``Entanglement Discrimination Catalysis''. Based on this set of states, we construct a new set $\mathrm{K}$ consisting of four locally indistinguishable states such that $\mathrm{K}^{\otimes m}$ (with $4^m$ members) is locally distinguishable for some $m$ greater than one. As an immediate application, we construct a noisy quantum channel with one sender and two receivers whose local zero-error classical capacity can achieve the full dimension of the input space but only with a multi-shot protocol.
\end{abstract}
\maketitle
\textit{Introduction.}---One of the main goals of quantum information theory is to understand the power and the limitation of quantum operations that can be implemented by local operations and classical communication (LOCC). These operations are natural requirements when two or more physically distant parties are trying to accomplish an information processing task. The class of LOCC operations has been playing a crucial role in a number of active researches in exploring the intrinsic properties of quantum information, especially in understanding the weird nature of quantum entanglement and quantum nonlocality.

One fundamental topic among these lines of research that has recently attracted lots of attention is the local distinguishability of quantum states. In the well studied bipartite case, a state secretly chosen from a set of pre-specified orthogonal quantum states is shared between two distant parties, say Alice and Bob. Their goal is to locally figure out the exact identity of this state \cite{NAT05,GKR+01,FAN04,OH06,BDFM+99,BDMS+99,GKRS04,BGK11,WSHV00,HMM+06,WAT05,YDY11,HSSH03,WH02,DFJY07,DFXY09}. In some special cases Alice and Bob are able to accomplish the discrimination without error and in many other cases they are not. For example, Walgate $et~al.$ proved that any two orthogonal pure states, no matter entangled or not, are locally perfectly distinguishable \cite{WSHV00}. Other interesting examples include sets of orthogonal pure states that are locally indistinguishable \cite{BDFM+99, BDMS+99}. Later, Horodecki $et~al.$ showed a phenomenon of ``more nonlocality with less entanglement''\cite{HSSH03}. These examples demonstrate that entanglement is not an essential feature of locally indistinguishable states. It is thus of great interest to study the role of entanglement in the local distinguishability problem. Considerable efforts have been devoted to the local discrimination of maximally entangled states. Ghosh $et~al.$ proved that any three Bell states cannot be discriminated with certainty by LOCC \cite{GKR+01}. In general, Alice and Bob are not able to locally distinguish $d+1$ or more $d\otimes d$ maximally entangled states with certainty \cite{GKR+01,NAT05,HMM+06,FAN04,OH06}. It would be quite interesting to know whether $d+1$ is always a tight lower bound for the number of locally indistinguishable maximally entangled states. In other words, whether there is any locally indistinguishable set consisting of $d$ maximally entangled states in $d\otimes d$ state space? This question was attacked in Refs. \cite{NAT05,HMM+06,FAN04,OH06,GKRS04,BGK11}, and the only known result is that in the case of $d=3$, any three orthogonal maximally entangled states are locally distinguishable \cite{NAT05}. It has been conjectured in Ref. \cite{BGK11} that for $d>3$ such set of states should exist by proposing four $4\otimes 4$ maximally entangled states that are locally indistinguishable by one-way LOCC. However, the possibility of distinguishability of these states by the most general LOCC operations has not been excluded.

In this Letter we resolve the above question by explicitly exhibiting four orthogonal ququad-ququad maximally entangled states that are not locally distinguishable. Our construction is remarkably simple, by accompanying each state from the standard Bell basis (i.e., $2\otimes 2$ maximally entangled states) with different Bell states. More precisely, our example is of the form $\{\ket{\Psi_0}\otimes \ket{\Psi_0}$, $\ket{\Psi_1}\otimes \ket{\Psi_1}$, $\ket{\Psi_2}\otimes \ket{\Psi_1}$, $\ket{\Psi_3}\otimes \ket{\Psi_1}\}$, where $\{\ket{\Psi_i}\}_{i=0}^3$ is the standard Bell basis. Based on the construction, we show how entanglement can be used, without being consumed, to accomplish state discrimination that cannot be achieved with certainty without it. In other words, with a $2\otimes 2$ maximally entangled state as resource, one can distinguish among the above four orthogonal ququad-ququad maximally entangled states, and after the discrimination, we are still left with another two-qubit maximally entangled state. This novel phenomenon is called ``Entanglement Discrimination Catalysis". It is worth noting that this phenomenon is different from the previously discovered catalysis by entanglement in the context of entanglement transformation \cite{JP99} and non-local quantum operations \cite{VC02}. Based on this phenomenon, we find a set $\mathrm{K}$ of four locally indistinguishable states such that $\mathrm{K}^{\otimes m}$ consisting of an exponentially large number of states $4^m$, is locally distinguishable. This indicates that local distinguishability of a set of states could be increased under tensor operation, a subtle fact previously overlooked. As an interesting application, we construct a noisy quantum channel with one classical sender and two quantum receivers whose local zero-error classical capacity can achieve the full dimension of the input space but only by using the channel multiple times. Intuitively, a noisy quantum channel could be boosted into a noiseless channel for sending classical information in the multi-shot scenario. The existence of such channel reveals a sharp difference between quantum channels with one receiver and those with two receivers. For quantum channel with one sender and one receiver, it was shown that entangled inputs cannot make imperfect quantum channels perfect \cite{BEHY10}, that is, for any such quantum channel, multi-shot can never render noisy quantum channels having maximum capacity, even asymptotically; if the sender is classical, Shor proved that the classical capacity of such channel is additive \cite{SHO02}.

The major difficulty in proving the local indistinguishability of the set of constructed states is that the structure of LOCC operations is mathematically complicated. We conquer this obstacle by showing that even a wider class of quantum operations that completely preserve the positivity of partial transpose (PPT) cannot distinguish these states. Since the set of LOCC operations is just a subset of that of PPT operations, local indistinguishability of these states by PPT operations immediately implies that of LOCC operations. Comparing to LOCC and Separable operations, PPT operations have a simpler mathematical structure that can be feasibly characterized by Semi-Definite Programming. One motivation of studying the state discrimination by PPT operations is its significant role in entanglement theory. In fact, PPT operations have been used to study the separability, entanglement distillation, and entanglement transformation \cite{PER96,HHH96,HHH98,RAI01,ISH04,MW08}. It was proved that PPT criterion is a necessary condition for the separability of quantum states \cite{PER96,HHH96}. Horodecki $et~al.$ showed that if a mixed state is distillable, it must violate the PPT criterion \cite{HHH98}. Ishizaka showed that bipartite pure entangled states can be transformed into another bipartite pure state with arbitrary high Schmidt rank by stochastic PPT operations \cite{ISH04}.

Before we present our main results, let us first review some notations and preliminaries. We shall use $\varphi$ to represent the density operator form $\op{\varphi}{\varphi}$ for a pure state $\ket{\varphi}$.  We also use $\ket{\Psi_i}$ to denote the standard Bell states with $\ket{\Psi_i}=(I_2\otimes \sigma_i)\frac{1}{\sqrt{2}}(\ket{00}+\ket{11})$, where $\sigma_i$s are the Pauli matrices given by $\sigma_0=I_2$ and
\begin{eqnarray*}
\sigma_1=\left(
\begin{array}{cc}
 1 & 0  \\
 0 & -1 \\
\end{array}
\right),
\sigma_2=\left(
\begin{array}{cc}
 0 & 1  \\
 1 & 0 \\
\end{array}
\right),
\sigma_3=\left(
\begin{array}{cc}
 0 & -i  \\
 i & 0 \\
\end{array}
\right).
\end{eqnarray*}

A Positive Operator-Valued Measure (POVM) with $n$ outcomes is an $n$-tuple of operators $(M_0,M_1,\cdots, M_{n-1})$ such that $M_i\geq 0$ and $\sum_{i=0}^{n-1} M_i=I$. A set of quantum states $\{\ket{\varphi_i}\}_{i=0}^{n-1}$ can be distinguished by POVM $(M_i)_{i=0}^{n-1}$ iff $M_i\ket{\varphi_i}=\ket{\varphi_i}$.
A general PPT discrimination is achieved by performing a PPT POVM where each element has positive partial transpose. More precisely, a POVM $(M_i)_{i=0}^{n-1}$ acting on a bipartite system $\mathcal{A}\otimes\mathcal{B}$ is said to be PPT if $M_i^{\Gamma_{\mathcal{A}}}\geq 0$ holds for $0\leq i\leq n-1$, where ${\Gamma_{\mathcal{A}}}$ means the partial transpose with respect to system  $\mathcal{A}$, i.e., $(\op{ij}{kl})^{\Gamma_{\mathcal{A}}}=\op{kj}{il}$. For simplicity, $\Gamma$ is used for $\Gamma_{\mathcal{A}}$ whenever it is clear from the context. It is known that the set of LOCC POVMs is a subset of the set of PPT POVMs. In other words, any POVM that can be realized by means of an LOCC protocol is also a PPT POVM.

Let $\mathcal{C}$, $\mathcal{D}$, and $\mathcal{F}$ be three POVMs with $n$ outcomes. Then $\mathcal{C}=W\mathcal{D}W^{\dag}$ for some matrix $W$ means that $C_i=WD_iW^{\dag}$ holds for any $0\leq i\leq n-1$. $\mathcal{F}=\lambda \mathcal{C}+(1-\lambda)\mathcal{D}$ denotes a convex combination of $\mathcal{C}$ and $\mathcal{D}$, i.e., $F_i=\lambda C_i+(1-\lambda)D_i$ holds for any $0\leq i\leq n-1$.

It is straightforward to verify the following useful properties concerning with the Bell diagonal bipartite operators.
\begin{prop}\label{prop1} Let $M$ be a linear operator over $2\otimes 2$ state space. Then 1). $\sum_{i=0}^3 (\sigma_i\otimes\sigma_i) M(\sigma_i\otimes\sigma_i)$ is diagonal under Bell basis; 2) If $M=\sum\limits \nu_i\Psi_i$ is diagonal in Bell basis, the partial transpose $M^{\Gamma}=\sum\limits \mu_i\Psi_i$ with $\mu_i=\mathrm{Tr}M/2-\nu_{4-i}$.  Thus $M,M^{\Gamma}\geq 0$ if and only if $0\leq2\nu_i\leq\mathrm{Tr}M$ for $0\leq i\leq 3$.
\end{prop}

\textit{Main Result.}---Let $\mathcal{A}$ and $\mathcal{B}$ both be $d$-dimensional Hilbert spaces  held by Alice and Bob, respectively. Then we can derive an upper bound on the number of PPT distinguishable orthogonal maximally entangled states.
\begin{theorem}
No $k>d$ maximally entangled states in $\mathcal{A}\otimes \mathcal{B}$ can be perfectly  distinguished by PPT operations.
\end{theorem}
\textbf{Proof:}-- We first show that if $E,E^{\Gamma}\geq 0$ and $E\ket{\Phi}=\ket{\Phi}$, then $\mathrm{tr} E\geq d$, where $\ket{\Phi}=\sum_{j=0}^{d-1}\ket{j}_{A}\ket{j}_{B}/\sqrt{d}$ is the standard maximally entangled state in space $\mathcal{A}\otimes \mathcal{B}$. Noticing that $\Phi^{\Gamma}\leq I/d$, we have
\begin{eqnarray*}
1=\mathrm{Tr}(\Phi)=\mathrm{Tr}(E\Phi)=\mathrm{Tr}(E^{\Gamma}\Phi^{\Gamma})\leq \mathrm{Tr}(E^{\Gamma}I/d)={\mathrm{Tr}(E)}/{d}.
\end{eqnarray*}
Now assume that a set of maximally entangled states $\{\ket{\Phi_i}\}_{i=0}^{k-1}$ can be distinguished by PPT POVM $(E_i)_{i=0}^{k-1}$. Then $\mathrm{tr}E_i\geq d$. The result of the theorem immediately follows from
\begin{eqnarray*}
d^2=\mathrm{Tr}I=\mathrm{Tr}(\sum_{i=0}^{k-1}E_i)=\sum_{i=0}^{k-1}\mathrm{Tr}E_i\geq kd \Rightarrow k\leq d.
\end{eqnarray*}
\hfill $\blacksquare$

The result in Theorem 1 is slightly stronger than the previous results where only separable operations were employed \cite{NAT05}. However, the key fact we have derived in the proof has been implicitly obtained in Refs. \cite{HMM+06,VT99}. The proof presented above seems new.

We shall now provide four ququad-ququad orthogonal PPT indistinguishable maximally entangled states which resolves the open problem mentioned in the introduction part. More precisely, we show that $\mathrm{S}=\{\ket{\chi_i}_{AB}:0\leq i\leq 3\}\subset\mathcal{A}\otimes \mathcal{B}$ cannot be distinguished by any PPT POVM with $\mathcal{A}=\mathcal{A}_0\otimes \mathcal{A}_1$ and $\mathcal{B}=\mathcal{B}_0\otimes \mathcal{B}_1$, where $\mathcal{A}_0,\mathcal{A}_1$, $\mathcal{B}_0$, $\mathcal{B}_1$ are all two-dimensional Hilbert spaces and
\begin{align*}
\ket{\chi_0}_{AB}=& ~\ket{\Psi_0}_{A_0B_0}\otimes\ket{\Psi_0}_{A_1B_1},\\
\ket{\chi_1}_{AB}=& ~\ket{\Psi_1}_{A_0B_0}\otimes\ket{\Psi_1}_{A_1B_1},\\
\ket{\chi_2}_{AB}=& ~\ket{\Psi_2}_{A_0B_0}\otimes\ket{\Psi_1}_{A_1B_1},\\
\ket{\chi_3}_{AB}=& ~\ket{\Psi_3}_{A_0B_0}\otimes\ket{\Psi_1}_{A_1B_1}.
\end{align*}
\begin{theorem}\label{main}
$\mathrm{S}$ cannot be distinguished perfectly by any PPT POVM.
\end{theorem}
\textbf{Proof:}---Let $\mathrm{M}_{\mathrm{S}}$ denote the set of PPT POVMs that can distinguish $\mathrm{S}$, i.e.,
\begin{eqnarray*}
\mathrm{M}_{\mathrm{S}}=\{(M_i)_{i=0}^3:M_i\ket{\chi_i}=\ket{\chi_i},\sum_{i=0}^3 M_i=I,M_i,M_i^{\Gamma}\geq 0\}.
\end{eqnarray*}
We shall show that $\mathrm{M}_{\mathrm{S}}$ is nonempty will lead to a contradiction.

The complete proof is rather complicated and lengthy. For ease of presentation, we shall outline the key proof ideas as follows, and leave some technical details in the supplementary material \cite{EPAPS}. By the nonempty assumption, we can choose $\mathcal{C}=(C_i)_{i=0}^3$ from $\mathrm{M}_{\mathrm{S}}$. One can then construct a new POVM $\mathcal{N}=(N_i)_{i=0}^3\in \mathrm{M}_{\mathrm{S}}$ with highly symmetrical properties by exploring the convexity and symmetries of $\mathrm{S}$. The form of $\mathcal{N}$ enables us to derive a contradiction by calculating its partial transpose directly to show that $\mathcal{N}$ can not distinguish $\mathrm{S}$, i.e., $\mathcal{N}\notin \mathrm{M}_{\mathrm{S}}$. Thus, one can conclude that $\mathrm{M}_{\mathrm{S}}$ has to be empty.

Now we start to describe how to construct the desired POVM $\mathcal{N}$. We need explore some properties of $\mathrm{M}_{\mathrm{S}}$ and $\mathrm{S}$.

Firstly, $\mathrm{M}_{\mathrm{S}}$ is convex, i.e., for any $0\leq\lambda\leq 1$,
$$\mathcal{C},\mathcal{D}\in \mathrm{M}_{\mathrm{S}}\Rightarrow \lambda \mathcal{C}+(1-\lambda)\mathcal{D}\in \mathrm{M}_{\mathrm{S}}.$$

Secondly, $\mathrm{S}$ enjoys a number of symmetries:\\
\textbf{S1}. For any Pauli matrix $\sigma$ and any $j=0,1$, $\sigma_{A_j}\otimes\sigma_{B_j}$ preserves $\ket{\chi_i}$ in the following way,
\begin{equation}
(\sigma_{A_j}\otimes\sigma_{B_j})\ket{\chi_i}=\pm\ket{\chi_i}.\nonumber
\end{equation}
\textbf{S2}. $W_{A_0B_0}$ preserves $\ket{\chi_0}$ and rotates $\ket{\chi_i}$ to $\ket{\chi_{i+1 \mod 3}}$ for $i=1,2,3$,
\begin{eqnarray}
W_{A_0B_0}\ket{\chi_0}=\ket{\chi_0},~~W_{A_0B_0}\ket{\chi_1}=\ket{\chi_2},\nonumber\\
W_{A_0B_0}\ket{\chi_2}=\ket{\chi_3},~~W_{A_0B_0}\ket{\chi_3}=\ket{\chi_1},\nonumber
\end{eqnarray}
where
$$W=\frac{1}{2}\left(
\begin{array}{cc}
 -i & 1  \\
 -i & -1 \\
\end{array}
\right)\otimes \left(
\begin{array}{cc}
 i & 1  \\
 i & -1 \\
\end{array}
\right).$$
\textbf{S3}. $U_{A_0B_0}$ preserves $\ket{\chi_0}$ and $\ket{\chi_1}$, and swaps between $\ket{\chi_2}$ and $\ket{\chi_3}$,
\begin{eqnarray}
U_{A_0B_0}\ket{\chi_0}=\ket{\chi_0},~U_{A_0B_0}\ket{\chi_1}=~~~\ket{\chi_1},\nonumber\\
U_{A_0B_0}\ket{\chi_2}=\ket{\chi_3},~U_{A_0B_0}\ket{\chi_3}=-\ket{\chi_2}.\nonumber
\end{eqnarray}
where
$$U=\left(
\begin{array}{cc}
 1 & 0  \\
 0 & -i \\
\end{array}
\right)\otimes \left(
\begin{array}{cc}
 1 & 0  \\
 0 & i \\
\end{array}
\right).$$
\textbf{S4}. For $\theta\in [0,2\pi)$,  $V(\theta)_{A_1B_1}$ preserves $\ket{\chi_i}$ for $0\leq i\leq 3$,
\begin{equation}
V(\theta)_{A_1B_1}\ket{\chi_i}=\ket{\chi_i}.\nonumber
\end{equation}
where
$$V(\theta)=\left(
\begin{array}{cc}
 1 & 0  \\
 0 & e^{-i\theta} \\
\end{array}
\right)\otimes \left(
\begin{array}{cc}
 1 & 0  \\
 0 & e^{i\theta} \\
\end{array}
\right).$$

Noticing that local unitary does not change the positivity of partial transpose, we can construct a POVM $\mathcal{N}=(N_i)_{i=0}^3\in \mathrm{M}_{\mathrm{S}}$  by the convexity of $\mathrm{M}_{\mathrm{S}}$ and \textbf{S1}-\textbf{S4} ($j=0,1$) such that
\begin{equation}\label{goal1}
N_1=U_{A_0B_0}N_1U_{A_0B_0}^{\dag},~N_{i+1}=W_{A_0B_0}N_iW_{A_0B_0}^{\dag}
\end{equation}
for $i=1,2$,
and
\begin{equation}\label{goal2}
\mathcal{N}=V(\theta)_{A_1B_1}\mathcal{N}V(\theta)_{A_1B_1}^{\dag}=(\sigma_{A_j}\otimes\sigma_{B_j})\mathcal{N}(\sigma_{A_j}\otimes\sigma_{B_j}).
\end{equation}
In particular, the second equality of Eq. (\ref{goal2}) indicates that the members of $\mathcal{N}$ are all diagonal in Bell basis, and Eq. (\ref{goal1}) has further greatly restricted the form of $\mathcal{N}$.

We shall obtain the required $\mathcal{N}$ from any POVM $\mathcal{C}\in \mathrm{M}_{\mathrm{S}}$ by the following four relatively simpler steps:

\textbf{Step 1}: Notice that for Pauli matrix $\sigma$, $$(\sigma_{A_0}\otimes\sigma_{B_0})\mathcal{C}(\sigma_{A_0}\otimes\sigma_{B_0})\in \mathrm{M}_{\mathrm{S}}.$$
Invoking \textbf{S1}, the convexity of $\mathrm{M}_{\mathrm{S}}$, and Proposition \ref{prop1}, we know that
$$\mathcal{D}=(\sum_{\sigma}(\sigma_{A_0}\otimes\sigma_{B_0})\mathcal{C}(\sigma_{A_0}\otimes\sigma_{B_0}))/4\in \mathrm{M}_{\mathrm{S}},$$
and each measurement operator $D_i$ is of the form $\sum_j \Psi_j\otimes D^{(ij)}$ for $0\leq i\leq 3$.

\textbf{Step 2}: According to \textbf{S2}, one can verify that
\begin{eqnarray*}
\mathcal{F}&=&W_{A_0B_0}(D_0,D_3,D_1,D_2)W_{A_0B_0}^{\dag}\in \mathrm{M}_{\mathrm{S}},\\
\mathcal{G}&=&W_{A_0B_0}^{\dag}(D_0,D_2,D_3,D_1)W_{A_0B_0}\in \mathrm{M}_{\mathrm{S}}
\end{eqnarray*}
Invoking the convexity of $\mathrm{M}_{\mathrm{S}}$ again, we have
\begin{equation*}
\mathcal{J}=(\mathcal{D}+\mathcal{F}+\mathcal{G})/3\in \mathrm{M}_{\mathrm{S}}.
\end{equation*}
Then we know that $J_0=W_{A_0B_0}J_0W_{A_0B_0}^{\dag}$ and for $i=1,2$,
\begin{equation*}
J_{i+1}=W_{A_0B_0}J_iW_{A_0B_0}^{\dag}.
\end{equation*}
\textbf{Step 3}: Define $\mathcal{K}=(K_i)_{i=0}^3$ such that
\begin{equation*}
K_0=J_0,~K_1=W_{A_0B_0}J_1W_{A_0B_0}^{\dag},~K_{i+1}=W_{A_0B_0}K_iW_{A_0B_0}^{\dag},
\end{equation*}
where $i=1,2$.
According to \textbf{S3}, we have $\mathcal{K}\in\mathrm{M}_{\mathrm{S}}$. Therefore,
\begin{equation*}
\mathcal{L}=(L_i)_{i=0}^3=(\mathcal{J}+\mathcal{K})/2\in \mathrm{M}_{\mathrm{S}}.
\end{equation*}
We know that for $i=1,2$,
\begin{equation*}
L_1=U_{A_0B_0}L_1U_{A_0B_0}^{\dag},~~L_{i+1}=W_{A_0B_0}L_iW_{A_0B_0}^{\dag}.
\end{equation*}
\textbf{Step 4}: Invoking \textbf{S4}, we obtain that
\begin{equation*}
\mathcal{L(\theta)}=(L_i(\theta))_{i=0}^3=V(\theta)_{A_1B1}\mathcal{L}V(\theta)_{A_1B1}^{\dag}\in \mathrm{M}_{\mathrm{S}},
\end{equation*}
then $\mathcal{M}=(\int\limits_{0}^{2\pi}L_i(\theta)d\theta)_{i=0}^3\in \mathrm{M}_{\mathrm{S}}$.
One can readily verify that $\mathcal{N}$ satisfies Eqs.(\ref{goal1}) and (\ref{goal2}), where
\begin{eqnarray*}
\mathcal{N}=(N_i)_{i=0}^3=(\sum_{\sigma}(\sigma_{A_0}\otimes\sigma_{B_0})\mathcal{M}(\sigma_{A_0}\otimes\sigma_{B_0}))/4\in \mathrm{M}_{\mathrm{S}}.
\end{eqnarray*}
The rest of the proof is to show that such $\mathcal{N}$ can not distinguish $\mathrm{S}$, i.e., $N_i,N_i^{\Gamma}\geq 0$, $N_i\ket{\chi_i}=\ket{\chi_i}$, $\sum N_i=I$ and Eqs. (\ref{goal1},\ref{goal2}) cannot be satisfied simultaneously. We refer the interested reader to the supplementary material for a detailed calculation \cite{EPAPS}.
\hfill $\blacksquare$

Since every LOCC POVM is also a PPT POVM, one can conclude that $\mathrm{S}$ is locally indistinguishable. To the best of our knowledge, this is the first example of $d$ orthogonal $d\otimes d$ maximally entangled states that are locally indistinguishable.

Due to the special structure of $\mathrm{S}$, we further observe a quite surprising ``Entanglement Discrimination Catalysis" phenomenon happens on $\mathrm{S}$. More precisely, with a two-qubit maximally entangled state as resource, says $|\Psi_0\rangle$, we can distinguish among the members of $\mathrm{S}$ locally, and after the discrimination, we are still left with an intact copy of $|\Psi_0\rangle$. The scheme is very simple: we use the given entanglement resource to distinguish the states of subsystem $\mathcal{A}_0\otimes \mathcal{B}_0$ via a teleportation protocol, then the outcome $i$ indicates that the original state to be distinguished is just $\ket{\chi_i}_{AB}$. After that, one can recover $|\Psi_0\rangle$ from the state of subsystem $\mathcal{A}_1\otimes \mathcal{B}_1$: If $i=0$, then the state of subsystem $\mathcal{A}_1\otimes \mathcal{B}_1$ is $\ket{\Psi_0}$. Otherwise, the state of subsystem $\mathcal{A}_1\otimes \mathcal{B}_1$ is $\ket{\Psi_1}$, and one can obtain an exact copy of $\ket{\Psi_0}$ by applying $\sigma_1$ to subsystem $\mathcal{B}_1$.

It is interesting that not only catalysts help local entanglement discrimination, but also tensor operation does. We shall show that there is some locally indistinguishable set $\mathrm{K}$ such that for some finite $m>1$, $\mathrm{K}^{\otimes m}$ becomes locally distinguishable, where the tensor product $\mathrm{S}_1\otimes \mathrm{S}_2$ of two sets $\mathrm{S}_1$ and $\mathrm{S}_2$ is defined as $\{\ket{s_1}\otimes{\ket{s_2}:\ket{s_i}\in \mathrm{S}_i}\}$.  Before presenting this set $\mathrm{K}$, we shall point out an interesting property of PPT distinguishability:  If a set of states is PPT indistinguishable, then sharing an entangled pure state with a sufficiently small amount of entanglement cannot make them PPT distinguishable. This property can be regarded as a direct consequence of the fact that the set of PPT POVMs with a fixed number of outcomes is a closed set. Here we provide a slightly refined statement. Suppose the optimal average success probability for distinguishing a set of bipartite orthogonal states $\{\rho_i:0\leq i\leq n-1\}$ on $\mathcal{A}\otimes \mathcal{B}$ with a priori probability distribution $\{p_0,\cdot\cdot\cdot,p_{n-1}\}$ by PPT POVM is $q$ where $q<1$. That is, $\sum\limits_{i=0}^{n-1}p_i \mathrm{tr}(E_i\rho_i)\leq q$ is valid for any PPT POVM $(E_i)_{i=0}^{n-1}$. Then we have
\begin{lemma}\label{2}
$\{\rho_i\otimes \alpha:0\leq i\leq n-1\}$ is PPT indistinguishable for any $\ket{\alpha}_{AB}=\sqrt{1-\varepsilon}\ket{00}+\sqrt{\varepsilon}\ket{11}$ with $0\leq \varepsilon<(1-q)^2$.
\end{lemma}
\textbf{Proof:}---For any PPT POVM $(E_i)_{i=0}^{n-1}$, the average success probability for distinguishing $\{\rho_i\otimes \alpha:0\leq i\leq n\}$ with a prior probability distribution $\{p_0,\cdot\cdot\cdot,p_{n-1}\}$ is
\begin{eqnarray*}
&&\sum_{i} p_i\mathrm{tr}(E_i(\rho_i\otimes \alpha))\\
&=&\sum_{i} (p_i\mathrm{tr}(E_i(\rho_i\otimes \beta))+p_i\mathrm{tr}(E_i(\rho_i\otimes(Q-S))))\\
&\leq& q+\sum_{i} p_i\mathrm{tr}(E_i(\rho_i\otimes Q))\\
&\leq& q+\sum_{i} p_i\mathrm{tr}(\rho_i\otimes Q)
= q+\mathrm{Tr}Q
= q+\sqrt{\varepsilon}<1,
\end{eqnarray*}
where $Q$ and $S$ are rank-$1$ positive operators with orthogonal support such that $\alpha-\beta=Q-S$ with $\beta=|00\rangle\langle 00|$. Thus, $\mathrm{tr}Q=\sqrt{\varepsilon}$.

One can therefore conclude that $\{\rho_i\otimes \alpha:0\leq i\leq n-1\}$ cannot be distinguished by PPT POVM. \hfill $\blacksquare$

According to Theorem \ref{main} and Lemma \ref{2}, we can choose a partially entangled state $\ket{\beta}_{AB}=\sqrt{1-\delta}\ket{00}+\sqrt{\delta}\ket{11}$ with $0<\delta< 1/2$, such that $\mathrm{K}=\mathrm{S}\otimes \{\ket{\beta}\}$ is PPT indistinguishable.
\begin{theorem}\label{main1}
There exists some finite $m$ such that $\mathrm{K}^{\otimes m}$ is locally distinguishable.
\end{theorem}
\textbf{Proof:}---We shall see that $\mathrm{K}^{\otimes m}=\mathrm{S}^{\otimes m}\otimes \{\ket{\beta}^{\otimes m}\}$ can be distinguished by the following two-step LOCC protocol, where $m=\lceil-\frac{1}{\log_2(1-\delta)}\rceil$:

Step 1: Transform $\ket{\beta}_{AB}^{\otimes m}$ into $\ket{\Psi_0}$ by LOCC, which can be accomplished according to the condition for entanglement transformation between bipartite pure states \cite{NIE99}.

Step 2: Use $\ket{\Psi_0}$ to distinguish $\mathrm{S}^{\otimes m}$: For any state  $\ket{\chi_{i_1}}\otimes\ket{\chi_{i_2}}\otimes\cdot\cdot\cdot\otimes\ket{\chi_{i_m}}\in \mathrm{S}^{\otimes m}$, by using $\ket{\Psi_0}$, we can identify $i_1$ and get another $\ket{\Psi_0}$, then identify $i_2$ and obtain $\ket{\Psi_0}$ again, etc. After identifying $i_1,i_2\cdot\cdot\cdot i_m$, the discrimination is finished.
\hfill $\blacksquare$

The above set $\mathrm{K}$ has an interesting implication in studying the local zero-error classical capacity of quantum channels, where LOCC discrimination has important applications \cite{WAT05,YDY11,DS08}. We can construct a channel $\mathcal{E}:\{1,2,3,4\}\mapsto \mathcal{B}\otimes\mathcal{C}$ with one classical sender Alice and two quantum receivers Bob and Charlie as follows:  For an input $0\leq i\leq 3$, the output $\ket{\chi_i}_{BC}\otimes\ket{\beta}_{BC}$ is distributed between Bob and Charlie. According to Theorem 2 and Lemma \ref{2}, Bob and Charlie are not able to distinguish the output states perfectly. Thus we know that the one-shot local zero-error classical capacity of $\mathcal{E}$ is strictly less than $\log_2 4=2$ bits. Suppose Alice now sends $i_1i_2\cdot\cdot\cdot i_m$, where $m$ is chosen as in the proof of Theorem \ref{main1}. We know that Bob and Charlie can identify $i_1i_2\cdot\cdot\cdot i_m$ perfectly. Thus Alice can transmit $\log_2 4^m=2m$ bits perfectly, which means that multi-shot of $\mathcal{E}$ can render this noisy quantum channel to have optimal capacity.

The techniques used in the proof of Theorem \ref{main} can be employed to study many other problems by PPT operations. For instance, we can show that $\sqrt{2/3}\ket{00}+\sqrt{1/3}\ket{11}$ is the minimal entanglement resource required for distinguishing three Bell states under PPT POVM. More precisely, we have
\begin{theorem}\label{main2}
$\mathrm{T}=\{\ket{\Psi_i}_{A_0B_0}\otimes \ket{\alpha}_{A_1B_1}\}_{i=1}^3$ is PPT distinguishable for a normalized $\ket{\alpha}=\sum\limits_{i=0}^{n-1}\sqrt{\lambda_i}\ket{ii}$ with $\lambda_0\geq \lambda_1\geq\cdot\cdot\cdot\geq\lambda_{n-1}\geq 0$ if and only if $\lambda_0\leq 2/3$.
\end{theorem}
A detailed proof of this theorem, together with other new results will be presented in a forthcoming paper.

\textit{Conclusion.}--- In this Letter, we present four orthogonal ququad-ququad maximally entangled states that cannot be distinguished locally. Later, the Phenomenon ``Entanglement Catalyst Discrimination'' is observed. Based on this result, we show that there is a set $\mathrm{K}$ which is locally indistinguishable, but $\mathrm{K}^{\otimes m}$ can be distinguished by LOCC for some $m>1$. Then we construct a classical-quantum noisy channel where multi-shot can make it noiseless in transmitting classical information.

${Acknowledgements.}$ We thank Professor C. H. Bennett for suggesting us to use ``Entanglement Catalyst Discrimination'' instead of ``Entanglement Discrimination Catalyst''. This work was partly supported by the National Natural Science Foundation of China
(Grant Nos. 61179030 and 60621062) and the Australian Research Council (Grant
Nos. DP110103473 and DP120103776).

\section{Supplementary material: the rest part of the proof of Theorem 2}
We only need to show that PPT POVM $\mathcal{N}$ cannot distinguish $\mathrm{S}$ if it enjoys that for $j=0,1$
\begin{eqnarray}\label{symetrical}
N_1=U_{A_0B_0}N_1U_{A_0B_0}^{\dag},~N_{i+1}=W_{A_0B_0}N_iW_{A_0B_0}^{\dag},{i=1,2},\\
\mathcal{N}=V(\theta)_{A_1B_1}\mathcal{N}V(\theta)_{A_1B_1}^{\dag}=(\sigma_{A_j}\otimes\sigma_{B_j})\mathcal{N}(\sigma_{A_j}\otimes\sigma_{B_j}).
\end{eqnarray}
To see this, first note that each $N_i$ has the following special form
\begin{eqnarray} \label{form}
N_i=\left(
\begin{array}{cccc}
 P_i & 0 & 0& T_i \\
 0 & R_i & 0 & 0 \\
 0 & 0 & R_i & 0\\
 T_i & 0 & 0 & P_i
\end{array}
\right),~~~N_i^{\Gamma}=\left(
\begin{array}{cccc}
 P_i^{\Gamma} & 0 & 0& 0 \\
 0 & R_i^{\Gamma} & T_i^{\Gamma} & 0 \\
 0 & T_i^{\Gamma} & R_i^{\Gamma} & 0\\
 0 & 0 & 0 & P_i^{\Gamma}
\end{array}
\right)
\end{eqnarray}
where $P_i$ $R_i$ and $T_i$ are Bell diagonal Hermitian on $\mathcal{A}_0\otimes \mathcal{B}_0$.

Furthermore, we can assume that
\begin{eqnarray*}
P_1&=&a_0\Psi_0+a_1\Psi_1+a_2\Psi_2+a_2\Psi_3,\\
T_1&=&b_0\Psi_0+b_1\Psi_1+b_2\Psi_2+b_2\Psi_3.
\end{eqnarray*}
According to Eq. (4), we have
\begin{eqnarray*}
P_0&=&(1-3a_{0})\Psi_0+(1-a_1-2a_2)(I-\Psi_0),\\
T_0&=&3a_{0}\Psi_0+(1-a_1-2a_2)(I-\Psi_0).
\end{eqnarray*}
$\mathcal{N}=(N_i)_{i=0}^3$ can distinguish $\mathrm{S}$ implies that
\begin{eqnarray*}
N_j\ket{\chi_j}=\ket{\chi_j}\Rightarrow a_1-b_1=1, a_2-b_2=0,a_0+b_0=0.
\end{eqnarray*}
$\mathcal{N}$ is a PPT POVM, then
\begin{eqnarray*}
N_1\geq 0\Rightarrow P_1\geq |T_1|&\Rightarrow& a_i\geq 0,|a_1-1|\leq a_1\\ &\Rightarrow& 1/2\leq a_1\leq 1,\\
N_0\geq 0\Rightarrow P_0\geq |T_0|&\Rightarrow& a_1+2a_2\leq 1,3a_0\leq (1-3a_0)\\ &\Rightarrow& 0\leq a_0\leq 1/6,\\
N_i^{\Gamma}\geq 0\Rightarrow{P_0}^{\Gamma}\geq 0&\Rightarrow& 1-3a_0\leq  3-3(a_1+2a_2),\\
N_i^{\Gamma}\geq 0\Rightarrow{P_1}^{\Gamma}\geq 0&\Rightarrow& a_1\leq a_0+2a_2,
\end{eqnarray*}
where $|A|=\sqrt{A^\dag A}$ denotes the positive square root of $A^{\dag}A$.\\
Applying the bound of $a_0,a_1,a_2$ obtained above, we see
$$0\leq 1-6a_0\leq 3-3(a_1+2a_2+a_0)\leq 3-3(a_1+a_1)\leq 0.$$
Now we can conclude that
\begin{eqnarray*}
&&a_1=1/2,~~a_2=a_0=1/6,\\
\Longrightarrow&&|T_i^{\Gamma}|=1/3(\Psi_0+\Psi_1+\Psi_2),~\mathrm{for}~0\leq i\leq 3.
\end{eqnarray*}
Therefore,
\begin{eqnarray*}
\sum\limits_{i=0}^3 N_i=I \Rightarrow I=\sum\limits_{i=0}^3 R_i^{\Gamma} ,\\
N_i^{\Gamma}\geq 0 \Rightarrow R_i^{\Gamma}\geq |T_i|,\\
\Longrightarrow I\geq \frac{4}{3}(\Psi_0+\Psi_1+\Psi_2).
\end{eqnarray*}
This is impossible. Thus $\mathcal{N}=(N_i)_{i=0}^3\notin \mathrm{M}_{\mathrm{S}}$, and the proof of Theorem 2 is complete. \hfill $\blacksquare$
\end{document}